\documentclass[12pt]{article}
\usepackage{amsmath}
\usepackage{amssymb,amsthm}
\usepackage{color}

\numberwithin{equation}{section}

\theoremstyle{definition}

\theoremstyle{remark}

\title{Landau superfluids as non equilibrium stationary states}
\author{Walter F. Wreszinski\\
        Instituto de Fisica USP\\
        Rua do Mat\~{a}o, s.n., Travessa R 187\\
        05508-090 S\~{a}o Paulo, Brazil\\
        \texttt{wreszins@gmail.com}
        }
\begin{document}
\maketitle
\begin{abstract}
We define a superfluid state to be a nonequilibrium stationary state (NESS), which, at zero temperature, satisfies certain \emph{metastability conditions},
which physically express that there should be a sufficiently small energy-momentum transfer between the particles of the 
fluid and the surroundings (e.g., pipe). It is shown that two models, the Girardeau model and the Huang-Yang-Luttinger (HYL) model 
describe superfluids in this sense and, moreover, that, in the case of the HYL model, the metastability condition is directly related to Nozi\`{e}res' 
conjecture that, due to the repulsive interaction, the condensate does not suffer fragmentation into two (or more) parts, 
thereby assuring its quantum coherence. The models are 
rigorous examples of NESS in which the system is not finite, but rather a many-body system.
\end{abstract}

\section{Introduction and Summary}

Superfluidity of a Bose fluid (e.g.~Helium IV) remains an outstanding and fascinating theoretical problem (see the 
complementary reviews \cite{Kad} and \cite{Leg1}, as well as the book \cite{Leg2}). In particular, Kadanoff in \cite{Kad} 
(see also \cite{BaPe}) recently rather sharply questioned the relevance of Landau's criterion (\cite{LaL},\cite{CDZ}, \cite{WSi}) to
the superfluid property. The latter may be roughly stated in the 
following way: by the flow of a fluid along a pipe, momentum may be lost to the walls, if the modulus of the velocity $|\vec{v}|$ is 
greater than
$$
v_{c} \equiv {\rm min}_{\vec{p}}\frac{\epsilon(\vec{p})}{|\vec{p}|} \, , 
\eqno{(1.1)}
$$
where $\epsilon(\vec{p})$ are the energies of the ``elementary excitations'' generated by friction.

This indicates that the concept of superfluidity still lacks a clear and precise theoretical foundation. 
One reason, argued in \cite{Kad}, Kadanoff, is that ``given the 
many mechanisms for broadening the distributions of both energy and momentum, it seems very implausible that a condition like (1.1) 
can begin to account for the very long-lived nature of the flow of superfluid Helium'' --- see also \cite{BaPe}. This argument is essentially 
supported by the results of the present paper (see remark 3). In the sequel, however, Kadanoff suggests \cite{Kad} that superfluidity is 
brought about by the existence of a coherent, ``macroscopic'' wave-function of type
$$
\Psi(\vec{x}) = \exp \left[\tfrac{i\chi(\vec{x})}{\hbar} \right]
\eqno{(1.2)}
$$
with $\chi$ possibly complex; see also the more complete analysis of (\cite{Leg1},\cite{Leg2}). This macroscopic wave-function is precisely 
the complex classical field occurring in the definition of ODLRO, and explains the two-fluid model and the London-Landau superfluid 
hydrodynamics (\cite{Lo},\cite{Kha}, \cite{La}; see also \cite{MaRo} for a nice textbook treatment, and \cite{SeW}, pg.~7, for a related discussion). 
Thus, ODLRO and the associated coherence properties of the condensate wave-function would suffice as the basis of the phenomenon of 
translational superfluidity, which would, then, not even require an interaction, at least conceptually, since the free Bose gas also exhibits 
ODLRO (see, e.g., \cite{MaRo}, pg.~119).

It seems to us, however, that a crucial element is missing in the above discussion, namely, the \emph{stability} of the condensate, a 
point brought up emphatically by Nozi\`{e}res in \cite{Noz}. Consider the situation in which one asks whether, instead of having at 
$T=0$ a condensate strictly in the lowest energy state, one could fragment it into two states $1$ and $2$, of arbitrarily close energies 
in the thermodynamic limit, with populations $N_{1}$ and $N_{2} = N-N_{1}$. Naturally, only the potential energy is able to distinguish 
between these two choices. Adopting a Hartree (mean-field) approximation with interaction Hamiltonian
$$
H_{N,V} = \frac{UN^{2}}{2V} \, , 
\eqno{(1.3)}
$$
with $U > 0$, he suggests that the energy costs $\frac{U[(N_{1}+N_{2})^{2}-N_{1}^{2}-N_{2}^{2}]}{2V} = \frac{UN_{1}N_{2}}{V}$, i.e., 
extensive in both $N_{1}$ and $N_{2}$ in the thermodynamic limit, where $U > 0$ is the strength of the interaction, assumed repulsive, 
and thus that it is the Coulomb interaction which interdicts the fragmentation, thereby assuring the condensate's quantum coherence, 
with the corresponding wave-function becoming a macroscopic observable. His argument in the above form is not directly relevant to 
superfluidity, because it is easy to see that the mean-field interaction Hamiltonian (1.3) does not exhibit superfluidity (in Landau's sense), 
see Section~3. We shall show in that section, however, that an effective Hamiltonian for dilute systems --- the Huang-Yang-Luttinger (HYL) 
model (\cite{HYL}, \cite{Hu}, \cite{BDLP}) --- does reproduce Nozi\`{e}res' heuristics in a precise sense.

The homogeneous Bose gas in a cubic box with periodic boundary conditions (b.c.) at zero temperature has also been recently studied in a 
paper by Cornean, Derezinski and Zin \cite{CDZ}. For the excitation spectrum of the weakly interacting Bose gas, see \cite{Sei}.

In order that the theory does not depend on the details of finite systems in a box, but only on quantities which remain fixed in the 
thermodynamic limit, we formulate our framework taking this into account in Section~2. We define a superfluid state to be a non-equilibrium stationary 
state (NESS) at zero temperature, satisfying 
certain \emph{metastability conditions}, which physically express that there should be a sufficiently small energy-momentum 
transfer between the particles of the fluid and the surroundings (e.g., the pipe). In Section~3 it is shown that two models, the Girardeau 
model \cite{Gi} and the Huang-Yang-Luttinger (HYL) model \cite{HYL} describe superfluids in this sense and, moreover, that, in the case 
of the HYL model, the metastability condition is directly related to the previously mentioned Nozi\`{e}res' conjecture.

The models are rigorous examples of NESS in which the system is not finite, 
but rather a many-body system. 

Our results do not rely on any assumptions on states of 
infinite systems.

\section{The general framework}

\subsection{General considerations}

The Landau condition (1.1) is supposed to lie at the bottom of a stability condition for the system in uniform motion with respect to a fixed, arbitrary inertial system. Although Landau formulated the latter in terms of ''elementary excitations'', we propose, along the lines of \cite{SeW}, to regard it in a preliminary step as a spectral condition imposed on a Hamiltonian describing a finite system. In order that this condition behaves smoothly when the thermodynamic limit is taken, it is necessary to consider, in some sense, the ''Hamiltonian of an infinite system'', and not just the thermodynamic limit of the ground state energy. A standard way of doing this is to consider the time dependent Green's functions \cite{Ba}. A fundamental requirement for this, in a rigorous approach, is to define an (algebra of) observables ''of the infinite system''.  

We consider Bosons in translational motion, to begin with in finite regions,which we take to be cubes, generically denoted by $\Lambda$.  The thermodynamic limit will be taken along the sequence of cubes
$$
\Lambda_{n}=[-nL,nL]^{d} \mbox{ of side } L_{n}=2nL \mbox{ with } n=1,2,3, \cdots 
\eqno{(2.1)}
$$
where $d$ is the space dimension. The number $L$ is arbitrary, and the sequence $\{n=1,2,3,\cdots \}$ may be replaced by any subset of the set of positive integers. Let 
$$
{\cal H}_{\Lambda_{j}} = L^{2}_{per}(\Lambda_{j}) \mbox{ for } j=1,2,3,\cdots
\eqno{(2.2.1)}
$$
denote the Hilbert space consisting of functions on $\mathbf{R}^{d}$, with $f\in L^{2}(\Lambda_{j})$ and such that f is periodic with period $L_{j}=2jL \mbox{ with } j=1,2,3,\cdots$ in each of the variables $\vec{x}=(x_{i}) \mbox{ with } i=1,\cdots,d$, i.e.,
$f(x_{i}+2jL)=f(x_{i}) \mbox{ with } 1=1,\cdots,d \mbox{ and } \vec{x}=(x_{i}), i=1,\cdots,d$. 
 Let
$$
{\cal F}_{\Lambda_{j}} \mbox{ denote the symmetrical Fock space over } {\cal H}_{\Lambda_{j}} \mbox{ with } j=1,2,3,\cdots
\eqno{(2.2.2)}
$$
We shall sometimes consider
$$
{\cal F}_{\Lambda_{j},N_{j}} = \otimes_{s,N} {\cal H}_{\Lambda_{j}} \mbox{ with } j=1,2,3,\cdots 
\eqno{(2.2.3)}
$$
, the symmetrized tensor product of  ${\cal H}_{\Lambda_{j}}$ corresponding to $N_{j}$ particles in $\Lambda_{j}$.
Our local algebras of observables will be taken as the (von Neumann) algebras of bounded operators on ${\cal F}_{\Lambda_{j}}$ (or on 
${\cal F}_{\Lambda_{j},N_{j}}$ where $N_{j}/|\Lambda_{j}|=\rho$, $\rho$ is the density and $|\Lambda|$ denotes the volume of $\Lambda$):
$$
{\cal A}_{\Lambda_{j}} =  {\cal B}({\cal F}_{\Lambda_{j}}) 
\eqno{(2.3)}
$$
By the choice (2.1) and (2.3) we have the isotony property
$$
{\cal A}_{\Lambda_{k}} \subset {\cal A}_{\Lambda_{l}} \mbox{ for } k<l \mbox{ or } \Lambda_{k} \subset \Lambda_{l}
\eqno{(2.4)}
$$
Indeed, a general element of ${\cal H}_{\Lambda_{k}}$ may be written $ f(\vec{x})=\sum_{\vec{n}\in \mathbf{Z}^{d}} c_{\vec{n}} \exp(2 \pi \vec{n} \cdot \vec{x}/L_{k})$ (or more precisely as a limit in the $L^{2}$ topology of finite sums), with $\sum_{\vec{n} \in \mathbf{Z}^{d}} |c_{\vec{n}}|^{2} < \infty$. Since functions with a given period are always periodic with a larger period, but not conversely, if $k<l$ or 
$\Lambda_{k} \subset \Lambda_{l}$, $f$ is also an element of ${\cal H}_{\Lambda_{l}}$, but, not conversely, so that the inclusion is strict. By the Riesz lemma, the same happens for the algebras ${\cal A}_{\Lambda_{k}}$. No changes in this argument arise on passing to ${\cal F}_{\Lambda_{j},N_{j}}$ or finally to ${\cal F}_{\Lambda_{j}}$ . 
Defining
$$ 
{\cal A}_{L} = \cup_{\Lambda_{k}} {\cal A}_{\Lambda_{k}} 
\eqno{(2.5.1)}
$$
as our local algebra, the norm-closed algebra 
$$
{\cal A} = \bar{{\cal A}_{L}} 
\eqno{(2.5.2)}
$$
is a C* algebra called the quasi-local algebra of the infinite system, where the bar denotes the C*-inductive limit \cite{KaRi}, Proposition 11.4.1): the isotony property (2.4) is crucial here. Since the local agebras are separable, because the Hilbert spaces ${\cal H}_{\Lambda_{j}}$ are separable, and the dual ${\cal B}({\cal H}_{\Lambda_{j}})$ is isomorphic to ${\cal H}_{\Lambda_{j}}$ by the Riesz lemma, the quasi-local algebra contains a countable (norm-) dense set (upon approximation by a suitable ${\cal A}_{\Lambda_{n}}$ for sufficiently large $n$). 

The necessity of this construction arises because for us it is imperative to define a momentum operator for the finite system, which requires some space translation invariance. Making the system into a torus is already a form of rending it ''infinite'' : the periodic b.c. are of global nature. For this reason, the isotony property does not hold in general, and a special choice  such as (2.1) had to be made. If, however, $L$ is irrational, the sequence 
$x_{n}=nL \mbox{ with } n=1,2,3,\cdots$  is uniformly distributed modulo 1 by Weyl's criterion (\cite{KN}, Example 2.1), so that the above special choice does not seem to imply any essential loss of generality.  

In $\Lambda_{j}$ (now generally simply called $\Lambda$) we consider a generic conservative system of $N$ identical particles of mass $m$. In units in which $\hbar = m = 1$, the generator of time translations, the Hamiltonian, will be denoted by $H_{\Lambda}$ and the generator of space translations, the momentum, will be denoted by $\vec{P}_{\Lambda}$ . As operators on ${\cal F}_{\Lambda}$ they take the standard forms
$$
H_{\Lambda} = \frac{-\sum_{r=1}^{N} \Delta_{r}}{2} + V(\vec{x}_{1}, \ldots , \vec{x}_{N})
\eqno{(2.6)}
$$
with $V$ a potential satisfying
$$
V \ge 0
\eqno{(2.7)}  
$$
i.e., only repulsive interactions will be considered, and
$$
\vec{P}_{\Lambda} = -i \sum_{r=1}^{N}\nabla_{r}
\eqno{(2.8)}
$$
with usual notations for the Laplacean $\Delta_{r}$ and the gradient $\nabla_{r}$ acting on the coordinates of the r-th particle.
We 
assume that $H_{\Lambda}$ and $\vec{P}_{\Lambda}$ are self-adjoint operators acting on 
${\cal F}_{\Lambda,N}$, with domains $D(H_{\Lambda})$ and 
$D(\vec{P}_{\Lambda})$, and 
$$
D(\vec{P}_{\Lambda}) \supset D(H_{\Lambda}) \, . 
\eqno{(2.9)}
$$
For ``sufficiently regular'' potentials in (2.6), (see \cite{MaRo}, and for precise conditions, (\cite{RSIV}, Chapter XIII.12 ), as well as for the models we shall treat, $H_{\Lambda}$ has a unique ground state (g.s) $\Omega_{\Lambda}$, i.e., 
$$
H_{\Lambda} \Omega_{\Lambda} = E_{\Lambda} \Omega_{\Lambda} \, , 
\eqno{(2.10.1)}
$$
where
$$
E_{\Lambda} \equiv \inf {\rm spec} (H_{\Lambda}) \, . 
\eqno{(2.10.2)}
$$
Unicity of the ground state and space-translation invariance of the Hamiltonian (2.6) imply that
$$
\vec{P}_{\Lambda} \Omega_{\Lambda} = \vec{0} \, . 
\eqno{(2.10.3)}
$$

We shall consider the so-called physical hamiltonian, normalizing the g.s. energy to zero, i.e.
$$
\widetilde{H}_{\Lambda} \equiv H_{\Lambda} - E_{\Lambda}
\eqno{(2.11.1)}
$$
which is therefore positive:
$$
\widetilde{H}_{\Lambda} \ge 0
\eqno{(2.11.2)}
$$
By thermodynamic stability,
$$
E_{\Lambda} \ge - c |\Lambda| \, , 
\eqno{(2.12)}
$$
where $|\Lambda|$ is the volume of $\Lambda$, and $c$ is a positive constant. In general, $E_{\Lambda}$ is of order of
$O(-d|\Lambda|)$ for some $d>0$, and in order to obtain a physical Hamiltonian satisfying (2.11.2) it is necessary to perform 
the renormalization (2.11.1) (infinite in the thermodynamic limit). For positive Hamiltonians with repulsive potential (2.7), this is also necessary,
otherwise the spectrum would tend to (plus) infinity in the thermodynamic limit. For instance, in the Girardeau model 
$E_{N,L} = \frac{(N-N^{-1})(\pi \rho)^{2}}{6}$ where $E_{N,L}$ denotes the ground state energy of $N$ particles in a periodic box of length $L$ and 
$\rho = \frac{N}{L}$ is the density. Note that the renormalization (2.11.1) is also physically imperative, because we shall be concerned with the spectrum of $H_{\Lambda}$ in a neighborhood of the ground state. Of course, the operators in the Heisenberg picture are not affected by the renormalization (2.11.1).

Let,now,
$$
S_{\Lambda}^{d} \equiv \left\{ \tfrac{2\pi\vec{n}}{L}\mid \vec{n} \in \mathbb{Z}^{d} \right\}
\eqno{(2.13.1)}
$$
and, given $\vec{v}_{lim} \in \mathbb{R}^{d}$, let 
$$
\vec{v}_{L}=\vec{v}_{\vec{n}_{L},L} = \vec{k}_{\vec{n}_{L},L}
\eqno{(2.13.2)}
$$ 
such that
\begin{eqnarray*}
|\vec{k}_{\vec{n}_{L},L}-\vec{v}| = \inf_{\vec{k} \in S_{\Lambda}^{d}}|\vec{k} - \vec{v}_{lim}| \\
\end{eqnarray*}$$\eqno{(2.13.3)}$$
If there is more than one $\vec{k}_{\vec{n}_{L},L}$ satisfying (2.13.3), we pick any one of them. We have:
$$
\lim_{N,L \to \infty}\vec{v}_{L} = \vec{v}_{lim} \, , 
\eqno{(2.14)}
$$
where $N,L \to \infty$ will be always taken to mean the thermodynamic limit, whereby
$$
N \to \infty \, , \quad L \to \infty \, , \quad \frac{N}{L^{d}} = \rho \quad \mbox{ with } 0<\rho<\infty \, , 
\eqno{(2.15)}
$$
where $\rho$ is a fixed density. The unitary operator of Galilei transformations appropriate to velocity 
$\vec{v}_{L}$ is given by
\begin{eqnarray*}
U_{\Lambda}^{\vec{v}} \equiv \exp \left(i \vec{v}_{L} \cdot (\vec{x}_{1} + \ldots + \vec{x}_{N}) \right) \, . 
\end{eqnarray*}
$$\eqno{(2.16)}$$
We shall assume (2.9) and that $U_{\Lambda}^{\vec{v}}$ maps $D(H_{\Lambda})$ into $D(H_{\Lambda})$. From now on we 
shall write $\vec{v}$ for $\vec{v}_{L}$ : the notation (2.14) assures that no confusion arises. It follows from (2.16), (2.6) and (2.8) that, on $D(H_{\Lambda})$,
\begin{eqnarray*}
(U_{\Lambda}^{\vec{v}})^{\dag} \widetilde{H}_{\Lambda} U_{\Lambda}^{\vec{v}}=\\ 
\widetilde{H}_{\Lambda} + \vec{v} \cdot \vec{P}_{\Lambda}+ \frac{N(\vec{v})^{2}}{2} \, , 
\end{eqnarray*}$$\eqno{(2.17)}$$
which is the expression of Galilean covariance. 
In addition to the standard model (2.6), we shall also consider effective Hamiltonians, for dilute Bose systems, which we define as follows. Assume that the limit
$$
e(\rho) \equiv \lim_{N,L \to \infty} L^{-d} E_{\Lambda} \, ,
\eqno{(2.18.1)} 
$$
exists, where $E_{\Lambda}$ is the ground state energy defined in (2.10.2), and $a$ denotes the scattering length corresponding 
to the potential $V$ in (2.6), which is also a measure of the interaction range; the mean particle distance is $\rho^{-1/3}$, 
now with $d=3$. It has been made plausible (see, e.g., \cite{Hu}, pg. 231) that at extremely low temperatures it is possible to describe a weakly interacting Bose gas only in terms of three parameters, $\lambda$, $\rho^{-1/3}$ and $a$, where $\lambda$ is the thermal wavelength. In the zero temperature limit we are reduced to $\rho^{-1/3}$ and $a$, and, if $a \ll \rho^{-1/3}$, or
$$
\rho a^{3} \ll 1
\eqno{(2.18.2)}
$$
we speak of a \textbf{dilute Bose gas}. It has been conjectured that $e(\rho)$ has an asymptotic expansion (see \cite{Yin} and references given there):
$$
e(\rho)= \sum_{j=1}^{k} e^{(j)}(\rho) + O((\rho)^{1/2})e^{(k)}(\rho) 
\eqno{(2.18.3)}
$$
as $\rho \to 0$, in $(\rho)^{1/2}$, with
$$
e^{1}(\rho) = 4\pi a \rho^{2} 
\eqno{(2.18.4)}
$$
and
$$ 
e^{2}(\rho) = e^{1}(\rho) \frac{128 (\rho a^{3})^{1/2}}{15\sqrt(\pi)} \, .
\eqno{(2.18.5)}
$$ 
We shall refer to the k-th approximant in (2.18.3), as approximant of order (k) and to $k$ as \textbf{order of the approximation}.
Several rigorous results support (2.18.4), it was rigorously proved by Lieb and Yngvason in a seminal paper \cite{LY} that 
$e(\rho) = 4\pi a (\rho)^{2} + o(a(\rho)^{2})$, and various rigorous additional results are given and reviewed in \cite{Yin}, to 
which we also refer for the references. The famous second order correction, (2.18.5), was first conjectured by Lee and Yang 
in 1957 \cite{LeYa}, on the basis of the pseudopotential approximation of Lee, Huang and Yang \cite{LeHuYa}, and the 
binary-collision expansion method \cite{Hu}.

\textbf{Definition 1} We call a Hamiltonian $H_{\Lambda}$ an \emph{effective Hamiltonian} to order $k$ for the dilute Boson 
system, if it satisfies both (2.18.6), (2.17) (Galilean covariance) and space-translation invariance.

In particular, if the ground state is unique, as in the HYL model, (2.10.3) also holds.

The article ''an'' in definition 2.1 expresses the fact that the definition does not specify a unique Hamiltonian but rather a \textbf{family} of Hamiltonians, which, unlike (2.6), depend explicitlt on $N,L$ and are therefore less fundamental (hence the name ''effective''), but do retain the basic physical symmetries (this latter fact is explicitly used in proposition 2 on the HYL model). We may therefore hope that effective Hamiltonians provide an approximate description of the physical properties of a dilute system.  

As we shall see, the Bogoliubov approximation does not define an effective Hamiltonian according to the above 
definition, because it lacks Galilean covariance. The pseudopotential approximation does define an 
effective Hamiltonian: for $j=1$, see Section~3.2, for $j=2$, see the conclusion.

We shall henceforth assume that the ground state $\Omega_{\Lambda}$ is unique.  

From (2.17) we may be led, as in \cite{SeW}, to ask whether the Hamiltonian
$$
H_{\vec{v},\Lambda} = \widetilde{H_{\Lambda}} + \vec{v} \cdot \vec{P}_{\Lambda} 
\eqno{(2.19.1)}
$$
is the one appropriate to describe the Bose fluid in uniform motion with velocity $\vec{v}$. By (2.17),
$$
H_{\vec{v},\Lambda} \ge -\Delta E_{\vec{v}}(\Lambda) = -\frac{N(\vec{v})^{2}}{2} \, . 
\eqno{(2.19.2)}
$$
On the other hand, by (2.17),
\begin{eqnarray*}
H_{\vec{v},\Lambda} = (U_{\Lambda}^{\vec{v}})^{\dag} \widetilde{H}_{\Lambda} U_{\Lambda}^{\vec{v}}\\
- \Delta E_{\vec{v}}(\Lambda) \, , 
\end{eqnarray*}$$\eqno{(2.19.3)}$$
from which
\begin{eqnarray*}
{\rm spec} \, (H_{\vec{v},\Lambda}) = {\rm spec} \, (\widetilde{H}_{\Lambda}) - \Delta E_{\vec{v}}(\Lambda)
\end{eqnarray*}$$\eqno{(2.19.4)}$$
follows. By (2.19.4), $-\Delta E_{\vec{v}}(\Lambda) \in {\rm spec} \, (\widetilde{H}_{\Lambda})$, which, together with (2.19.2), implies that
$$
-\Delta E_{\vec{v}}(\Lambda) = \inf {\rm spec} \, (H_{\vec{v},\Lambda}) \, . 
\eqno{(2.20)}
$$

Incidently, this proves that the Bogoliubov approximation (BA) is not Galilean covariant: this is a consequence of the 
fact that, in the BA, the operator $H_{\vec{v},\Lambda}$ is non-negative for $|\vec{v}|$ sufficiently small (see \cite{ZBru}), 
contradicting (2.20), which was derived on the basis of Galilean covariance. 

By (2.20), one might be led to consider the vector $\Psi_{\vec{v},\Lambda}$ corresponding to the lowest eigenvalue (2.20), 
and the corresponding state $\langle \Psi_{\vec{v},\Lambda}, \cdot \,  \Psi_{\vec{v},\Lambda} \rangle$, with the renormalization
$H_{\vec{v},\Lambda} \to \widetilde{H}_{\vec{v},\Lambda} + \Delta E_{\vec{v}}(\Lambda) \ge 0$ as describing the ground state of the 
Bose fluid in motion with velocity $\vec{v}$, which would yield an equilibrium state in the thermodynamic limit
$N,L \to \infty$. We have, however, the following result. 

\textbf{Lemma 1}
$U_{\Lambda}^{-\vec{v}}|\Omega_{\Lambda} \rangle$ is the unique eigenvector of $H_{\vec{v},\Lambda}$ corresponding to 
the eigenvalue $-\Delta E_{\vec{v}}(\Lambda)$.

\textbf{Proof} 
Apply (2.19.3) to $(U_{\Lambda}^{\vec{v}})^{\dag}|\Omega_{\Lambda})= U_{\Lambda}^{-\vec{v}}|\Omega_{\Lambda})$. 

Lemma 1 shows that $\Psi_{\vec{v},\Lambda} = U_{\Lambda}^{-\vec{v}}|\Omega_{\Lambda}\rangle$ represents a state of the Bose fluid in 
uniform motion with velocity $-\vec{v}$. Our aim is, however, to look at the system, initially in an equilibrium (ground) state,
$|\Omega_{\Lambda})$, when set into uniform motion with velocity $\vec{v}$, not $-\vec{v}$, or, alternatively, to look at it from the point of view of a moving observer with the opposite velocity $-\vec{v}$ : by (2.17), (2.11.1) and (2.10.3), its initial energy will be the kinetic energy $\Delta E_{\vec{v}}(\Lambda)$ associated to this velocity field. According to Landau, in order to detect friction in this system, one should compare this quantity with the eigenvalues (or ''elementary excitations'') of the operator $ U_{\Lambda}^{\vec{v}})^{\dag} \widetilde{H}_{\Lambda} U_{\Lambda}^{\vec{v}}$, which, again by by (2.17) and definition (2.19.1), ammounts to analysing the spectrum of $H_{\vec{v},\Lambda}$: if it is positive for a given range of values of $\vec{v}$, no dissipation occurs.

Again by (2.10.3) and (2.11.1), $|\Omega_{\Lambda})$  is an eigenstate of $H_{\vec{v},\Lambda}$ of eigenvalue zero, so that stationarity is guaranteed.

The ground state of $H_{\vec{v},\Lambda}$ corresponds, by Lemma 1, to the reversion of the velocities 
of a macroscopic number of particles. This explains the connection to metastability: very large momentum is necessary to connect the original ground state to this state. This fact explains, at the same time, the conceptual necessity of considering the thermodynamic limit. Equations (2.19.3) and (2.20) suggest that, ``in normal circumstances'', $H_{\vec{v},\Lambda}$ has a spectrum contained 
in $[-\Delta E_{\vec{v}}(\Lambda),\infty)$, which tends to the whole real line as $N,L \to \infty$. In order to see what may go wrong, assume 
that the spectrum ${\rm spec} \, (H_{\vec{v},\Lambda}) = [-\Delta E_{\vec{v}}(\Lambda), - \lambda_{N,L}]\cup[0,\infty)$, with $\lambda_{N,L} \to \infty$ 
as $N,L \to \infty$. In this case, the negative part of the spectrum ``disappears'' in the thermodynamic limit, and the assertion that the spectrum 
of the physical hamiltonian (for the infinite system) contains a non-empty set in the negative real axis does not follow. What we need is a certain 
uniformity in the thermodynamic limit. This is our next subject.
     
\subsection{The thermodynamic limit}

We begin by defining
\begin{eqnarray*}
\alpha_{\Lambda,t}(A) \equiv \exp(it \widetilde{H}_{\Lambda}) A \exp(-it \widetilde{H}_{\Lambda}) \\
\sigma_{\Lambda,\vec{x}}(A) \equiv \exp(i\vec{x}\cdot\vec{P}_{\Lambda}) A \exp(-i\vec{x}\cdot\vec{P}_{\Lambda})
\end{eqnarray*}$$\eqno{(2.21)}$$

The basic quantity is the Green's function for the finite system, defined as
\begin{eqnarray*}
G_{\Lambda}(A,B;t,\vec{x})\equiv \omega_{\Lambda}(A \alpha_{\Lambda,t}(\sigma_{\Lambda,\vec{x}}(B))\\
\forall A,B \in {\cal A}_{L} \mbox{ and } t \in \mathbf{R}
\end{eqnarray*}$$\eqno{(2.22.1)}$$

By the definition (2.5.2) of the quasi-local algebra ${\cal A}$, and (2.22.1), $G_{\Lambda}(A,B;t,\vec{x})$ may be extended to $A,B \in {\cal A}$, and the extension satisfies
$$
|G_{\Lambda}(A,B;t,\vec{x})| \le ||A||||B||
\eqno{(2.22.2)}
$$
By the choice (2.5.1) the ${\cal A}_{\Lambda_{n}}$ are separable, and thus, by the Banach-Alaoglu theorem (\cite{RSI}, Theorem IV.21), there exist subsequences (\cite{RSI}, Chapter IV) $\Lambda_{n_{k}}$ such that the limits 
\begin{eqnarray*}
G(A,B;t,\vec{x}) = \lim_{k} G_{\Lambda_{n_{k}}}(A,B;t,\vec{x}) \\
\mbox{ exist } \forall A,B \in {\cal A} \forall t \in \mathbf{R} \forall \vec{x} \in \mathbf{R}^{d}
\end{eqnarray*}
$$\eqno{(2.22.3)}$$

Thus a sequence $\Lambda_{n_{k}}$ such that the thermodynamic limit (2.22.3) 
exists can always be found but is not unique: such nonuniqueness is associated with phase transitions, and 
is thus expected, e.g., for the Bose fluid at sufficiently low temperatures.

Clearly $\omega$, defined by $$\omega(A) = G(A,\mathbf{1};0.\vec{0}) \eqno{(2.22.4)}$$ is a state over ${\cal A}$: a
state of the system is described by a linear functional $\omega$ on ${\cal A}$ which associates to each $A \in {\cal A}$ 
a number $\omega(A) \in \mathbb{C}$ such that $\omega(A^{*}A)) \ge 0$ for all $A \in {\cal A}$, and $\omega(\mathbf{1}) = 1$. By the GNS 
construction $\omega$ is induced by a cyclic vector $\Omega_{\omega}$, a (physical) separable Hilbert space ${\cal H}_{\omega}$, and 
a representation $\pi_{\omega}({\cal A})$ of ${\cal A}$ by bounded operators on ${\cal H}_{\omega}$, such that 
$\omega(A) = \langle \Omega_{\omega}, \pi_{\omega}(A) \Omega_{\omega}\rangle$ and $\overline{\pi_{\omega}({\cal A}) \Omega_{\omega}} 
= {\cal H}_{\omega}$. The following assumption (which we shall \textbf{not} make), ammounts to the requirement of existence of generators, which is, indeed,  a mild requirement, almost universally assumed to be a characteristic of the infinite system:

\textbf{Assumption A} 
\begin{eqnarray*}
G(A,B;t,\vec{x}) = \\
(\Omega_{\omega},\pi_{\omega}(A)\exp[i(tH_{\omega}+\vec{x} \cdot \vec{P}_{\omega})]\pi_{\omega}(B)\Omega_{\omega})
\end{eqnarray*}$$\eqno{(2.22.4)}$$
where $\Omega_{\omega}$ and $\pi_{\omega}$ are the the GNS vector and representation associated to $\omega$, and $H_{\omega}$ and $\vec{P}_{\omega}$ are the commuting (in the sense of spectral projections) self-adjoint generators of time and space translations in the representation $\pi$.

In case assumption A holds, (2.10.1), (2.10.3), (2.11.1), (2.22.1), (2.22.3), and (2.22.4) imply that
$$
\omega(\exp[i(tH_{\omega}+\vec{x} \cdot \vec{P}_{\omega})] A \exp[-i(tH_{\omega}+\vec{x} \cdot \vec{P}_{\omega})]) = \omega(A)
\eqno{(2.22.5)}
$$
i.e., $\omega$ is a time- and space- translation invariant state. In correspondence with (2.11.2), we define

\textbf{Definition 2} $\omega$ is a ground state of the infinite system if it is time-translation invariant and the generator $H_{\omega}$ of time-translations satisfies
$$
H_{\omega} \ge 0
\eqno{(2.23)}
$$

Above, only time translations are considered, but our g.s. will also be space-translation invariant. It may be proved (\cite{Hug}, theorem 3.4) that the ground state condition (2.23) is equivalent to an analyticity property of the Green's function (see \cite{Hug}, theorem 3.2), which may alternatively be taken as definition of the infinite-volume g.s.. 

For continuous Boson systems, however, even Assumption A is not easy to prove: see \cite{BRo2}, second edition, theorem 6.3.27 (ii). The $t-$ continuity of $G(A,B;t,\vec{x})$ was shown for a Bose gas with repulsive interactions in \cite{BraRo}, but only for strictly negative chemical potential. We are, however, interested in the case of a low-temperature Bose fluid, with BEC and adopt, therefore, a different strategy. We now also adopt our dynamics as defined by the $H_{\vec{v},\Lambda}$ defined by (2.19.1). Accordingly, we have the Green's function for the finite system
\begin{eqnarray*}
G_{\Lambda,\vec{v}}(A,B;t,\vec{x})\equiv \omega_{\Lambda}(A \alpha_{\Lambda,\vec{v},t}(\sigma_{\Lambda,\vec{x}}(B))\\
\forall A,B \in {\cal A}_{L} \mbox{ and } t \in \mathbf{R}
\end{eqnarray*}$$\eqno{(2.24)}$$
with
$$
\alpha_{\Lambda,\vec{v},t}(A) \equiv \exp(it H_{\vec{v},\Lambda}) A \exp(-it H_{\vec{v},\Lambda}) 
\eqno{(2.21')}
$$
together with the analogues of (2.22.2) and (2.22.3); the analogue of (2.22.4) is the same, of course: we are dealing with the same state under two different dynamics. The generator of time translations in the infinite-volume representation is, now denoted by $H_{\vec{v},\omega}$.

We now come back to the explicit notation (2.1) et seq. and define the \emph{energy-momentum spectrum} $emsp_{N_{k},L_{k}}$ of the finite system as 
the set of pairs $(E_{r,N_{k},L_{k}}(\vec{v}), \vec{k}_{s,L_{k}})_{r,s}$, where $\{E_{r,N_{k},L{k}}(\vec{v})\}_{r}$ denotes the set of eigenvalues of $H_{\vec{v},\Lambda_{k}}= H_{\vec{v},N_{k},L_{k}}$, and 
$\{\vec{k}_{s,L_{k}}\}_{s} = S_{\Lambda_{k}}^{d}$ (given by (2.13.1)) the set of eigenvalues of the momentum $\vec{P}_{N_{k},L_{k}}$. Above, when $\vec{v}$ occurs in the context of the finite system, (2.13.2), (2.13.3) is understood.

\textbf{Definition 3.1} $emsp_{\infty}^{l}(\vec{v}_{lim})$ is the set of limit points of $emsp_{N_{k},L_{k}}$, possibly along subsequences.

The $\vec{v}_{lim}$ in definition 3.1 above (recall (2.14)) is symbolic for the fact that the limit of a point in  $emsp_{N_{k},L_{k}}$ is performed taking (2.13.2), (2.13.3) into account. 

We recall (see, e.g., \cite{BB}, pg. 40) that a distribution $T$ on $\mathbf{R}^{d}$ is said to vanish on an open subset $\Omega \subset \mathbf{R}^{d}$ iff $T(\phi) = 0$ for all $\phi \in {\cal D}(\Omega)$ (see \cite{BB} for the standard definition of ${\cal D}(\Omega)$, as well as of the Schwartz spaces of tempered distributions ${\cal S}^{'}(\mathbf{R}^{d})$). The \textbf{support} $supp T$ of $T \in {\cal S}^{'}(\mathbf{R}^{d})$ is the complement of the largest open set $\Omega \subset \mathbf{R}^{d}$ on which $T$ vanishes. It is characterized by the formula
$$
supp T = \cap_{A \in C_{T}} A
$$
where $C_{T}$ denotes the set of all closed subsets of $\Omega$ such that $T$ vanishes on $\Omega - A$. Accordingly:

\textbf{Lemma 2} $x \in \Omega$ belongs to $supp T$ with $T \in {\cal S}^{'}(\mathbf{R}^{d})$ on $\Omega$ iff $T$ does not vanish in every open neighbourhood $U$ of $x$, i.e., for every open neighborhood $U$ of $x$ there is a $\phi \in {\cal D}(U)$ such that $T(\phi) \ne 0$.

Note that the restriction of a $T \in {\cal S}^{'}(\mathbf{R}^{d})$ to a nonempty subset $\Omega \subset \mathbf{R}^{d}$ is defined in \cite{BB}, pg.40.

When $emsp_{\infty}^{l}(\vec{v}_{lim}) \ne \emptyset$ (we shall prove this in certain examples), we have:  

\textbf{Theorem 1} 
\begin{eqnarray*}
emsp_{\infty}(\vec{v}_{lim}) \equiv \mbox{ The support of the Fourier transform of } \\
G_{\vec{v}}(A^{*},A;t,\vec{x}) \mbox{ as a tempered distribution }
\end{eqnarray*}$$\eqno{(2.25.1)}$$
where $G_{\vec{v}}(A^{*},A;t,\vec{x})$ is defined in (2.24), agrees with the support of the joint spectral family $E(\lambda,\vec{k})$ of $(H_{\vec{v},\omega},\vec{P}_{\omega})$  in case assumption $A_{\vec{v}}$ holds true , which means: with the l.h.s of (2.22.4) replaced by $G_{\vec{v}}(A,B;t,\vec{x})$, with $H_{\vec{v},\omega}$ as generator of time translations and $\vec{P}_{\omega}$ as generator of space translations. Furthermore:
$$
emsp_{\infty}^{l}(\vec{v}_{lim}) \subset emsp_{\infty}(\vec{v}_{lim}) 
\eqno{(2.25.2)}
$$
\textbf{Remark} Both in $H_{\vec{v},\omega}$ and $G_{\vec{v}}(A,B;t,\vec{x})$, $\vec{v}$ is symbolic for the limit (2.14); we do not assume, in Assumption
$A_{\vec{v}}$, that $(H_{\vec{v},\omega} = H_{\omega} + \vec{v} \cdot \vec{P}_{\omega}$, although this is conjectured to be so, because it is not needed.

\textbf{Proof} We have that
 $G_{\vec{v}}(A^{*},A;t,\vec{x})$ is the pointwise limit of a subsequence
\begin{eqnarray*}
G_{N_{n_{k}},L_{n_{k}}, \vec{v}}(A^{*},A;t,\vec{x}) = \\
(\Omega_{N_{n_{k}},L_{n_{k}}},A^{*}\exp(it(\widetilde{H}_{N_{n_{k}},L_{n_{k}}}+\\
+\vec{v}_{n_{n_{k}},L_{n_{k}}} \cdot \vec{P}_{N_{n_{k}},L_{n_{k}}}))\\   
\exp(i\vec{x} \cdot \vec{P}_{N_{n_{k}},L_{n_{k}}}) A \Omega_{N_{n_{k}},L_{n_{k}}}) 
\end{eqnarray*}
which is uniformly bounded
$$
|G_{N_{n_{k}},L_{n_{k}},\vec{v}}(A^{*},A;t,\vec{x})| \le \| A\| ^{2} \, . 
$$
Hence, the $G_{N_{n_{k}},L_{n_{k}},\vec{v}}$ converge as $k \to \infty$ to $G_{\vec{v}}$ as tempered distributions, and thus the Fourier transforms $\widehat{G}_{N_{n_{k}},L_{n_{k}},\vec{v}}$ also converge to $\widehat{G_{\vec{v}}}$ as tempered distributions.
In case assumption A holds, we have 
\begin{eqnarray*}
G_{\vec{v}}(A^{*},A;t,\vec{x})=\\ 
=(\pi_{\omega}(A)\Omega_{\omega} , {\rm e}^{itH_{\vec{v},\omega}}\\
{\rm e}^{i\vec{x}\cdot\vec{P}_{\omega}}\pi_{\omega}(A)\Omega_{\omega})\\
=\int{\rm e}^{it\lambda(\vec{k})}{\rm e}^{i\vec{x}\cdot\vec{k}}{\rm d}\|E(\lambda,\vec{k})\pi_{\omega}(A)\Omega_{\omega}\| ^{2}\\\forall  t \in \mathbb{R} \,,  \quad  \forall \vec{x} \in \mathbb{R}^{d} \, .
\end{eqnarray*}
where
$\{E(\lambda)E(\vec{k})\equiv E(\lambda,\vec{k})\}$ denote the spectral family associated to 
$(H_{\vec{v},\omega},\vec{P}_{\omega})$, i.e.,
$(H_{\vec{v},\omega},\vec{P}_{\omega}) = \int (\lambda,\vec{k}) dE(\lambda,\vec{k})$.
The latter is the energy-momentum 
spectrum of $(H_{\vec{v},\omega},\vec{P}_{\omega})$ . This proves the first assertion of theorem 1. To prove the second assertion, assume that $(E(\vec{k}_{0}), \vec{k}_{0}) = \lim_{k \to \infty} (E_{r,N_{n_{k}},L_{n_{k}}},k_{s,L_{n_{k}}})$ for given fixed $(r,s)$, i.e., that $(E(\vec{k}_{0}),\vec{k}_{0}) \in emsp_{\infty}^{l}(\vec{v})_{lim})$. We now use again the convergence of $\widehat{G}_{N_{n_{k}},L_{n_{k}},\vec{v}}$  to $\widehat{G_{\vec{v}}}$ as tempered distributions and the fact that the finite system has a discrete spectrum. Thus $\widehat{G}_{N_{n_{k}},L_{n_{k}},\vec{v}}$ consists of point measures at points arbitrarily close to $(E(\vec{k}_{0}),\vec{k}_{0})$. Therefore, if $\phi \in {\cal D}(U)$ with $U$ an arbitrary neighborhood of $(E(\vec{k}_{0}),\vec{k}_{0})$, is chosen as having a local, nonzero maximum at $(E(\vec{k}_{0}),\vec{k}_{0})$, we have that $G_{\vec{v}}(\phi) > 0$. By lemma 2 and (2.25.1), (2.25.2) follows. q.e.d.      

Theorem 1 justifies the

\textbf{Definition 3.2} $emsp_{\infty}(\vec{v})$ will be called the energy-momentum spectrum of the infinite system. If 
$\exists (\lambda,\vec{k}) \in emsp_{\infty}(\vec{v})$ such that $\lambda < 0$, we shall say that the system describes a NESS.

The last statement of definition 3.2 is justified by the fact that, when generators exist, i.e., assumption A holds, a zero-temperature state is a NESS iff $spec(H_{\vec{v},\omega}) \cap (-\infty,0) \ne \emptyset$ by definition 2.

\textbf{Corollary 1}
The zero temperature state obtained from (2.24) and (2.22.4)  is a NESS if $emsp_{\infty}^{l}(\vec{v})$ contains a point $(\lambda(\vec{k_{0}}),\vec{k_{0}})$ with $\lambda(\vec{k}_{0}) < 0$.

Searching, now, for a metastability condition, i.e., a condition which might account for the long-livedness of the superfluid state, we are led to define subspaces of states ''close'' to the superfluid state, restricted to which $H_{\vec{v},\Lambda}$ is expected to be positive.

One proposal, specially appropriate when BEC occurs, may be formulated in terms of the number operators for the different momenta. Let 
$N^{op} = \sum_{\vec{k} \in S_{N,L}^{d}} n_{\vec{k}}$ where $N^{op}$ is the number operator
$$
N^{op} = \int_{\Lambda}d\vec{x} \; \Psi^{\dag}(\vec{x}) \Psi(\vec{x})
\eqno{(2.26)}
$$
and $\Psi(f)$ are the smeared basic destruction operators satisfying
$$
[\Psi(f),\Psi(g)^{\dag}] = (f,g) 
\eqno{(2.27)}
$$
In terms of the basic destruction operator $\Psi$ the group of Galilei transformations is given by
$$
\Psi(\vec{x},t) \to \Psi(\vec{x}-\vec{v}t,t) \exp[i(\vec{v}\cdot\vec{x}+(\vec{v})^{2}t/2)]
\eqno{(2.28)}
$$
(the mass $m=1$). As explained, e.g., in \cite{WSi}, we may restrict the transformation (2.29) to $t=0$, in which case
$$
\Psi(\vec{x}) \to \Psi(\vec{x}) {\rm e}^{i\vec{v}\cdot \vec{x}} \, . 
\eqno{(2.29)}
$$
whose first quantized version is (2.17). The $n_{\vec{k}}$  is the number operator corresponding to momentum 
$\vec{k}$; we have $(\Omega_{\Lambda}, N^{op} \Omega_{\Lambda})=N$, where $\Omega_{\Lambda}$ denotes as in (2.10.1) the ground state of $H_{\Lambda}$ with zero momentum (2.10.3). As before, we often replace $\Lambda$ by $(N,L)$. Define
$$
(\Omega_{\Lambda}, \frac{n_{\vec{0}}}{L^{3}} \Omega_{\Lambda}) \equiv \rho_{\vec{0},\Lambda}
\eqno{(2.30.1)}
$$
and consider the subspace of ${\cal H}_{\Lambda}$ defined by
\begin{eqnarray*}
{\cal E}_{max}^{\Lambda} \equiv \{\Psi_{\Lambda};(\Psi_{\Lambda},(N - n_{\vec{0}}) \Psi_{\Lambda}) \\
\le [\rho_{max} L^{3}] (\Psi_{\Lambda},\Psi_{\Lambda})
\end{eqnarray*}
$$\eqno{(2.30.2)}$$
for a given $\rho_{max}$ independent of $N,L$, where $[x]$ denotes the greatest integer $\le x$. When BEC takes place, 
$$
\lim_{N,L \to \infty} \rho_{\vec{0},\Lambda} = \rho_{\vec{0}} >0
\eqno{(2.30.3)}
$$
where $\rho_{\vec{0}}$ has the interpretation of the density of condensate particles, and (2.30.2) means that ${\cal E}_{max}^{\Lambda}$ consists of sates ''close'' to $\Omega_{\Lambda}$ in the sense that, in these states, the density of particles of nonzero momentum is bounded by a critical constant independent of $N,L$. By expressing $\Psi_{\Lambda}$ in the ''occupation-number representation'' basis (denoting the eigenvalues of the opertators $n_{\vec{k}}$ by the same symbols) $|n_{\vec{0}}, \cdots, n_{\vec{k}}, \cdots )$ of ${\cal H}_{N,L}$, we see that ${\cal E}_{max}^{\Lambda}$ defines a subspace of ${\cal H}_{N,L}$, i.e., the spectral subspace of $\sum_{\vec{k} \ne \vec{0}} n_{\vec{k}}$ corresponding to eigenvalues $\sum_{\vec{k} \ne \vec{0}} n_{\vec{k}} \le [\rho_{max} L^{3}]$.
If $\rho_{\vec{0}}$ in (2.30.3) exists and equals to zero, for sufficiently large $N,L$ the r.h.s of (2.30.2) equals $\rho$, $\rho_{max}=\rho$, and the subspace becomes the whole space.

In general, $n_{\vec{k}}$ do not commute with $H_{\Lambda}$, but $\frac{n_{\vec{0}}}{L^{3}}$ may be replaced by a c-number in the thermodynamic limit in a precise sense (see \cite{LSYng} and references given there). In the example of section 3.2 the $n_{\vec{k}}$ do commute with $H_{\Lambda}$: this feature is, however, highly unrealistic, even in the framework of effective Hamiltonians: already in second order of the pseudopotential approximation this does not occur (see \cite{Hu}, pg. 330).

Another proposal, connected to Kadanoff's remark in the introduction related to the broadness in energy and momentum, is:
\begin{eqnarray*}
{\cal H}_{N,L}^{c,d} = \mbox{ the subspace of } {\cal H}_{N,L} \mbox{ generated by of all linear combinations of vectors }\\
\sum_{i,j} \lambda_{i,j} \Psi_{i,j} \mbox{ with } \Psi_{i,j} \mbox{ simultaneous eigenvectors of } \widetilde{H}_{N,L} \mbox{ and } P_{N,L}\\
\mbox{ corresponding to eigenvalues } E_{i,N,L}(\vec{v}=\vec{0}) \mbox{ and }\vec{k}_{j,L}\\
\mbox{ such that } E_{i,N,L} \le c \forall i \mbox{ and } |\vec{k}_{j,L}| \le d
\end{eqnarray*}$$\eqno{(2.31)}$$

It is important that in (2.31) $c,d$ are independent of $N,L$.

The restrictions in (2.31) on the magnitude of the energy-momentum around the ground state are due to the fact that we wish to express that there should be a sufficiently small energy-momentum transfer between the particles of the fluid and the surroundings (pipe). As remarked by Baym \cite{Ba}, 
footnote on pg.~132, there is implicit the assumption that the energy of the excitations (in our case: the spectrum of 
$H_{\vec{v}}$) does not depend on the velocity of the walls, an assumption which should be valid, but only for slow relative motion 
of the superfluid and walls. Accordingly, we pose:

\textbf{Definition 4} \, \qquad
The system defines a \emph{superfluid} at $T=0$ if $\exists 0<v_{c}<\infty$ such that, whenever $|\vec{v}| \le v_{c}$, one of the following holds \, \qquad
a.) there exists a subspace ${\cal R}_{N,L}^{c,d}$ of 
${\cal H}_{N,L}^{c,d}$ such that the eigenvalues $\{E_{j,N,L}(\vec{v})^{'}\}_{j}$ of the restriction of
$\widetilde{H}_{N,L} + \vec{v}_{\vec{n}_{L},L} \cdot \vec{P}_{N,L} = H_{\vec{v},N,L}$ to
${\cal R}_{N,L}^{c,d}$ satisfy
$$
\epsilon^{c,d}_{j}(\vec{v}) \equiv \lim_{N,L \to \infty} E_{j,N,L}(\vec{v}) \ge 0 \forall j
\eqno{(2.32.1)}
$$
and that
$$
\{\epsilon^{c,d}_{j}(\vec{v})\}_{j} \ne \{0\}
\eqno{(2.32.2)}
$$
for some $0<c<\infty$, $0<d<\infty$; \, \qquad
b.) the eigenvalues $\{E_{j,N,L}(\vec{v})^{''}\}$ of the restriction of
$H_{\vec{v},N,L}$ to the subspace defined by (2.30.2) satisfy
$$
\liminf_{N,L \to \infty} E_{j,N,L}(\vec{v})^{''} \ge 0
\eqno{(2.33)}
$$
We refer to (2.32) and (2.33) as \emph{metastability conditions}.\, \qquad

Definition 4 is related to Conjecture 2.2 of \cite{CDZ} and to \cite{Sei}, Corollary 1 and remarks thereafter.

For the moment, the reader should ignore the necessity of  further restricting  ${\cal H}_{N,L}^{c,d}$ to 
${\cal R}_{N,L}^{c,d}$, which is technical and related to the fact that ''elementary excitations'' are defined ''up to a $O(N^{-1}$ correction'': this will be discussed in the example of section 3.1, but is included in the definition because we believe it is a general feature of ''elementary excitations''.

A basic role in the above definition is played by the $(N,L)$- independence of $c.d$ in (2.31) and $\rho_{max}$ in (2.30.2); (2.32.2) guarantees that the subspace ${\cal R}^{c,d}_{N,L}$ does not shrink to the empty set as $N,L \to \infty$. \, \qquad
In case assumption A holds, we may define 
\begin{eqnarray*}
{\cal H}_{\omega}^{c,d} \equiv \int_{\lambda \le c,|\vec{k} \le d}dE(\lambda,\vec{k}) \;  {\cal H}_{\omega} \, .
\end{eqnarray*}$$\eqno{(2.34)}$$
and are led to the alternative (in case a.)):

\textbf{Definition 5}\, \qquad
The state $\omega$ defines a \emph{superfluid} at $T=0$ if 
$\exists 0<v_{c}<\infty$ such that either: there exists a subspace ${\cal H}_{\omega}^{c,d,'}$ of 
${\cal H}_{\omega}^{c,d}$ such that
$$
H_{\vec{v},\omega} \mbox{ when restricted to }{\cal H}_{\omega}^{c,d,'} \ge 0
\eqno{(2.35)}
$$
for some $0<c<\infty$, $0<d<\infty$.

The metastability conditions comprise a precise way of characterizing ``stability under creation of a few elementary excitations'', 
and, when they are satisfied, the model of independent quasiparticles introduced in \cite{SeW} may be sometimes shown to be 
effective, as we shall demonstrate in the case of the first model treated in Section~3.
   
\section{Examples}

We apply the theory of Section~2 to two examples; in each of them, the system is a NESS in a sense stronger than that stated 
in Theorem 1 : the energy spectrum will be shown to be unbounded from below.

In both subsections 3.1 and 3.2 it is understood that $L=L_{n}$ as in (2.1), $\vec{v}=\vec{v}_{L_{n}}$ in accordance with (2.13.1), (2.13.2) and (2.13.3), with $n$ sufficiently large. Because of (2.14), it is possible to state some conditions as (3.12) and (3.36) in propositions 1 and 2, respectively, in terms of $\vec{v}_{lim}$.

\subsection{The Girardeau model}

We start with the Lieb-Liniger model \cite{LLi} of $N$ particles (for simplicity odd) in one dimension with repulsive delta function 
interactions, whose formal Hamiltonian is given by
$$
H_{N,L} = -\sum_{i=1}^{N}\frac{\partial^{2}}{2 (\partial x)^{2}} + 2c (\sum_{i,j=1}^{N})^{'} \delta(x_{i}-x_{j}) 
\mbox{ with } 0 \le x_{i},x_{j} \le L  \, . 
\eqno{(3.1)}
$$
The prime over the second sum indicates that the sum is confined to nearest neighbors. For the (standard) rigorous definition 
corresponding to(3.1), see \cite{Do}.

The limit, as $ c \to \infty$, of equation (3.1), yields Girardeau's model \cite{Gi}. It is the free particle Hamiltonian with Dirichlet b.c.~on 
the lines $x_{i}=x_{j}$, with quadratic form domain given in (\cite{Do}, pg.~353, (2.18)-(2.19)). It is straightforward that $U_{\Lambda}^{\vec{v}}$, 
given by (2.16), leaves this domain invariant and (2.9) holds. It is essential that we speak of the finite system, for which $H_{\vec{v},N,L}$ is bounded from below, and quadratic form methods apply; note that (2.9) is strict, because $\vec{P}_{\Lambda}$ is unbounded from below also for the finite system.
 Our considerations should also apply to the much richer Lieb-Liniger model, 
but, since details are much simpler in the case of the Girardeau model, the essence of the following argument becomes clearer and more concise.

For $c \to \infty$, the b.c. on the wave-functions reduces to
$$
\Psi(x_{1},\ldots,x_{N}) = 0 \mbox{ if } x_{j}=x_{i}, \; 1 \le x_{i},x_{j} \le N \, , 
\eqno{(3.2)}
$$
and the (Bose) eigenfunctions $\Psi^{B}$ satisfying equation (3.2) simplify to
$$
\Psi^{B}(x_{1},\ldots,x_{N}) = \Psi^{F}(x_{1},\ldots,x_{N})A(x_{1},\ldots,x_{N}) \, , 
\eqno{(3.3)}
$$
where $\Psi^{F}$ is the Fermi wave-function for the free system of $N$ particles confined to the region 
$0 \le x_{i} < L,i=1,\ldots,N$, with periodic b.c., and
$$
A(x_{1},\ldots,x_{N}) = \prod_{j<l} sign(x_{j}-x_{l}) \, . 
\eqno{(3.4)}
$$
Note that $\Psi^{F}$ automatically satisfies equation (3.2) by the exclusion principle. Indicating the Bose and Fermi ground states by the 
subscript $0$, it follows from equation (3.3) and the non-negativity of $\Psi_{0}^{B}$ that
$$
\Psi_{0}^{B} = |\Psi_{0}^{F}| \, . 
$$
Since $A^{2} = 1$, by equation (3.4), the correspondence between $\Psi^{B}$ and $\Psi^{F}$ given by equation (3.3) preserves all scalar 
products, and therefore the energy spectrum of the Bose system is the same as that of the free Fermi gas. Furthermore, $\Psi^{F}$ is a 
Slater determinant of plane-wave functions labelled by wave vectors $k_{i},i=1,\ldots,N$, equally spaced over the range $[-k_{F},k_{F}]$,
where $k_{F}$ is the Fermi momentum. Hence $k_{F}=\pi\frac{N-1}{L}$, which, in the thermodynamic limit reduces to 
$$
k_{F} = \pi \rho \, .
\eqno{(3.5)} 
$$
The simplest excitation is obtained by moving a particle from $k_{F}$ to $q > k_{F}$ (or from $-k_{F}$ to $q < -k_{F}$) thereby leaving 
a hole at $k_{F}$ (or $-k_{F}$). This excitation has momentum $k=q-k_{F}$ (or $-(q-k_{F})$), and energy
$\epsilon(k)=(k^{2}-k_{F}^{2})/2$, i.e.,
$$
\epsilon^{1}(k)=k^{2}/2 + k_{F}|k| \, . 
\eqno{(3.6)}
$$
This type of excitation must be supplemented by the umklapp excitations, which we consider in a more general form than \cite{Lieb}. 
They consist in taking a particle from $(-k_{F}-p)$ to $(k_{F}+q)$ or $(k_{F}-q)$ to $(-k_{F}-p)$. We call the corresponding eigenvalues 
$\epsilon^{3}(k)$ and $\epsilon^{2}(k)$ and consider just the latter in detail; for these
$$
0 \le q \le 2\pi(N-1) \mbox{ and } \frac{2\pi}{L} \le p
\eqno{(3.7)}
$$
and their  momentum
$$
k=-2k_{F}-(p-q)
\eqno{(3.8)}
$$
and their energies $\epsilon^{2}(k)$ are given by
$$
2\epsilon^{2}(k) = [(-k_{F}-p)^{2}-(k_{F}-q)^{2}]=[2k_{F}+(p-q)](p+q) \, . 
\eqno{(3.9)}
$$
A different, equivalent choice involving holes may be made \cite{Lieb}. The complete set of eigenvalues of 
$\widetilde{H}_{N,L} = H_{N,L} - E_{N,L}^{0}$ is
\begin{eqnarray*}
E_{n_{i}m_{i}p_{i}}^{m,\vec{P}}(\vec{k},\vec{l},\vec{q})=\sum_{n_{i}, m_{i}, p_{i}=0,1;\sum_{i}(n_{i}+m_{i}+p_{i})=m}\\
( n_{i}\epsilon^{1}(\vec{k}_{i})+\\ 
+m_{i}\epsilon^{2}(\vec{l}_{i})+p_{i}\epsilon^{3}(\vec{q}_{i})) \,  
\end{eqnarray*}$$\eqno{(3.10)}$$
with $m=1,\ldots,N$ and $\vec{k}_{i} \in \Lambda_{*}=\{\frac{2\pi n}{L}\mid n \in \mathbb{Z}^{*}\} $ and $ \vec{l}_{i} $ belongs to the set specified by (3.7) and (3.8). In (3.10), the $\epsilon^{3}(\vec{q}_{i})$ are the umklapp excitations we left out: the momenta $\vec{q}_{i}$ are similar to (3.7) and (3.8) but need not be specified here.

In (3.10), $\epsilon^{1}$ and $\epsilon^{2}$ have different meanings: $\epsilon^{1}$ are elementary excitations as the sum 
involving $\epsilon^{1}(\vec{k}_{i})$ in (3.10) is only an eigenvalue of $\widetilde{H}_{N,L}$ up to $O(N^{-1})$; for the umklapp 
excitations, the same would follow if we considered, as in \cite{Lieb}, just the first one in (3.7)-(3.9), i.e., corresponding 
to $q=0,p=\frac{2\pi}{L}$. By considering the general case (3.7)-(3.9), we have no corrections ($O(N^{-1})$ or otherwise) 
as far as the energies of the umklapp excitations are concerned. The momenta are, however, strictly additive in all 
cases \cite{Lieb}: for $p_{i}=0$ the momentum $\vec{P}$ associated to $E_{n_{i},m_{i}}^{m,\vec{P}}(\vec{k},\vec{l})$ in (3.10) is
$$
\vec{P} = \vec{k} + \vec{l} \, \mbox{ where }  
\vec{k} = \sum_{i=1}^{r}\vec{k}_{i} \mbox{ and } 
\vec{l} = \sum_{j=1}^{s} \vec{l}_{j} \, , 
\eqno{(3.11)}
$$
with $r,s$ integers. We keep the vector index in the momenta to emphasize the (crucial) point that there are, of course, two directions, 
along the positive and negative axes. Without loss of generality we assume that $\vec{v}$ is directed along the positive axis, and denote $v=|\vec{v}|$. 
When momenta appear without a vector sign, their sign will define their orientation. With this choice, our restriction to the second type of umklapp excitations is clear: they have momentum opposite to $\vec{v}$ and thus lead to lower eigenvalues of $H_{v,N,L}$.

\textbf{Proposition 1} For the Girardeau model, if
$$
0<v_{lim}<2\pi\rho \, , 
\eqno{(3.12)}
$$
the spectrum ${\rm spec} \, (H_{v,N,L})$ of $H_{v,N,L}$ contains the set ${\cal E}$, defined by
$$
{\cal E} = \{-2k_{F}vj + o(L)\}_{j=1}^{m}
\mbox{ with } m=[\tfrac{Lv}{2\pi}] 
\eqno{(3.13)}
$$
Thus, the quantities
$$
\lambda_{j} \equiv -2k_{F}v j \mbox{ with } j=1,2, \ldots
\eqno{(3.14)}
$$
belong to $emsp_{\infty}^{l}$  in the sense of definition 3.1, which is thus  unbounded from below, and thus the system describes a NESS by corollary 1. Furthermore, if
$$
c= 2(\pi \rho)^{2} 
\eqno{(3.15)}
$$
and
$$
d=\pi \rho \, ,
\eqno{(3.16)}
$$
then the subspace (2.31) satisfies (2.32.1) and (2.32.2), and thus the system describes a superfluid at $T=0$.

The idea of the proof is very simple. In (3.10) we consider only a sufficiently large number of umklapp excitations; their momentum is given by (3.8) and thus opposite to $\vec{v}$ by our (arbitrary) choice (otherwise we would have taken the other set of umklapp excitations $\epsilon^{3}$ in (3.10)). Consider, now
$$
H_{v,N,L} = H_{N,L} + v P_{N,L}
\eqno{(3.17.1)}
$$
We now switch one particle from $k_{F}$ to $-k_{F}-p$ (corresponding to $q=0$, $p=\frac{2\pi}{L}$ in (3.8)). This corresponding elementary excitation aquires thereby the momentum 
$$
k_{1} = -2k_{F} - \frac{2\pi}{L}
\eqno{(3.17.2)}
$$
Taking now a particle from $k_{F}-\frac{2\pi}{L}$ to $-k_{F}-2\frac{2\pi}{L}$ (corresponding to $q=\frac{2\pi}{L}$, $p=2\frac{2\pi}{L}$ in (3.8)), this again leads to the same elementary excitation momentum (3.17.2), and the resulting momentum of the two excitations is $-2(2k_{F})+o(L)$. Proceeding successively in this manner, when the m-th particle is extracted we arrive at momentum $-2k_{F}m +o(L)$, with $-2k_{F}mv+o(L)$ being the eigenvalue of $vP_{N,L}$ in (3.17.1). The energy of $\tilde{H_{N,L}}$ is of order $O(m^{2}/L$ by (3.9). For the eigenvalues of the full $H_{v,N,L}$ in (3.17.1) we are thus led to search for the minimum of $f(m)=m^{2}/L - 2k_{F}mv$, roughly, of order $m=[Lv]$ for $L$ sufficiently large. On the other hand, this same argument leads to
$$\sum_{j=1}^{m} \epsilon^{2}(k_{j}) - \sum_{j=1}^{m+1} \epsilon^{2}(k_{j}) = -2k_{F}v + o(L)$$, which is (3.13).

\textbf{Proof} We take (3.10), $n_{i}=0, m_{i}=1$. Choose now $P= \sum_{j=1}^{s} l_{j}$ in (3.11), with, by (3.8), 
$l_{j}= -2k_{F}-(p_{j}-q_{j})$ and 
$$
p_{j} = \frac{2\pi j}{L} \mbox{ and } q_{j} = \frac{2\pi (j-1)}{L} \mbox{ with } j=1, \cdots, m
$$
obtaining for the eigenvalues of $H_{v,N,L}$, given by (3.17.1),
\begin{eqnarray*}
\sum_{j=1}^{m} \epsilon^{2}(l_{j}) = \frac{2k_{F}m(m+1)\pi}{L} - \frac{2k_{F}\pi m}{L} + \\
 \qquad + \frac{2(\pi)^{2}m(m-1)}{L^{2}} - 2k_{F} m v - \frac{2\pi m v}{L} \, . 
\end{eqnarray*}
$$\eqno{(3.18.1)}$$
The minimum of the right hand side of (3.18.1) coincides approximately (for large enough $L$) with that of the function
$$
f(m) = \alpha m^{2} - \beta m 
\quad \mbox{with} \quad
\alpha = \frac{2 \pi^{2}\rho}{L} \quad 
\mbox{and} \quad \beta = 2\pi \rho v \, , 
\eqno{(3.18.2)}
$$
which is situated at $m=[\frac{\beta}{2\alpha}]=[\frac{Lv}{2\pi}]$, for $L$ large enough, and is given by
$\frac{-(\beta)^{2}}{4\alpha}=\frac{-N(v)^{2}}{2}$; we have $[\frac{Lv}{2\pi}] \sim \frac{Lv}{2\pi}-1 \le N$ or
$v \le \frac{2\pi(N+1)}{L}$, which is satisfied under assumption (3.12). By (3.18.1), for $m=1$, we obtain for the r.h.s. 
$-2k_{F}v + o(L)$, and for the l.h.s. $\epsilon^{2}(l_{1})$. For general $m$ (3.18.1) also yields
$$
\sum_{j=1}^{m} \epsilon^{2} (l_{j}) - \sum_{j=1}^{m+1}\epsilon^{2}(k_{j}) = -2k_{F}v + o(L) \, . 
\eqno{(3.18.3)}
$$
Thus, the points of the set ${\cal E}$ are $\{\epsilon_{j,L}=-2k_{F}v j + o(L)\}_{j=1}^{m}$ with 
$-2k_{F}vm = \frac{-Nv^{2}}{2} + o(L)$, and (3.13) follows; therefore, 
$(\lambda(k_{0}),k_{0})$, with $\lambda(k_{0}) = -2k_{F}v j; k_{0}=-2k_{F} j$,  is a limit point of the sequence
$(\lambda_{N,L}^{j},k_{N,L}^{j}=-2k_{F}(j+\frac{\pi}{k_{F}L})$ as $N,L \to \infty$. By Corollary 1 the assertion connected to (3.14) follows.

Let, now, $c,d$ be as in (3.15), (3.16). If we restrict $|\vec{P}|$ in (3.10) to satisfy 
$$
|\vec{P}| < \pi \rho 
\eqno{(3.19.1)}
$$
and
$$
|\vec{k}| < \pi \rho
\eqno{(3.19.2)}
$$
then
$$
|\vec{l}| = |\vec{P}-\vec{k}| < \pi \rho + \pi \rho = 2\pi \rho \, , 
\eqno{(3.20)}
$$
contradicting (3.8). Thus, in (3.10), the umklapp excitations are absent.

We now show that (3.19.2) is satisfied when $H_{N,L}$ and $\vec{P}_{N,L}$ are restricted to the subspace 
\begin{eqnarray*}
{\cal R}^{c,d}_{N,L} \equiv \mbox{ subspace of } {\cal H}_{N,L}^{c,d}\\
\mbox{ corresponding to a fixed } r \mbox{ independent of } N,L\\
\mbox{ of elementary excitations } \epsilon^{1}(\cdot) \mbox{ in (3.10) }
\end{eqnarray*}$$\eqno{(3.21)}$$
Under the assumption
$$
H_{N,L}-E_{N,L}^{0} \mbox{ restricted to } {\cal R}^{c,d}_{N,L} \le c
\eqno{(3.22)}
$$
we have that, since 
$\epsilon^{1}(\vec{k}) \ge 0$, and $\epsilon^{2}(\vec{l}) \ge 0$, $\epsilon^{3}(\vec{q}) \ge 0$, (3.6), (3.10) and (3.22) imply 
$$
2\pi \rho (|\vec{k}_{1}|+\ldots+|\vec{k}_{r}|) \le c \mbox{ for } \vec{k}=\vec{k}_{1}+\ldots+\vec{k}_{r}
\eqno{(3.23.1)}
$$
for $N$ sufficiently large and fixed $r$ due to the $O(N^{-1})$ corrections, hence
\begin{align*}
|\vec{k}| & = |\vec{k}_{1}+ \ldots +\vec{k}_{r}| \le |\vec{k}_{1}|+ \ldots +|\vec{k}_{r}|\\
& \le \frac{c}{2 \pi \rho}=\pi\rho
\end{align*}$$\eqno{(3.23.2)}$$
by (3.15), proving (3.19.2).

In order to prove (2.32.1), we consider $H_{v,N,L}$ when restricted to the subspace defined by (3.21) where, by (3.19),
$$
|\vec{P}_{N,L}| \le d =\pi \rho \, . 
\eqno{(3.23.3)}
$$
By (3.10), (3.19.4) and what was shown above, the spectrum of $H_{\vec{v},N,L}$ restricted to ${\cal R}^{c,d}_{N,L}$  
consists of eigenvalues
\begin{eqnarray*}
\epsilon^{c,d}_{r,N,L} \equiv \{\sum_{i=1}^{r}(\epsilon^{1}(\vec{k}_{i}) + \vec{v}.\vec{k}_{i})\\ 
\mbox{ with }|\vec{k}_{1}+ \cdots + \vec{k}_{r}| \le d ; \sum_{i=1}^{r} \epsilon^{1}(\vec{k}_{i}) \le c \}
\end{eqnarray*}$$\eqno{(3.23.4)}$$
up to corrections $O(N^{-1})$ when $r$ is a fixed number (independent of $N$ and~$L$). By (3.6) and (3.11),
$\epsilon^{1}(\vec{k}_{i}) + \vec{v} \cdot \vec{k}_{i} \ge 0 $ for all $i=1,\ldots,r$, and thus, by (3.23.4), (2.32.1) holds. Moreover, defining $\epsilon^{c,d}_{r}(\vec{v}) \equiv \lim_{N,L \to \infty} \epsilon^{c,d}_{r,N,L}$ we have that
$\{\epsilon^{c,d}_{r}(\vec{v})\}_{r} \ne \{0\} $ for any $r \ge 1$ by (3.23.4) ,so that (2.32.2) also holds. q.e.d. \, \qquad

\textbf{Remark 1}
In case assumption A holds, defining by (2.34)
$$
{\cal H}_{\omega}^{c,d,'} = \int dE(\lambda,\vec{k}) \; {\cal H}_{\omega}^{c,d}
\eqno{(3.24.1)}
$$
with
\begin{eqnarray*}
(\lambda,\vec{k}) = (\epsilon^{1}(\vec{k}_{1})+\vec{v}\cdot \vec{k}_{1}+\\
\ldots +\epsilon^{1}(\vec{k}_{r})+\vec{v} \cdot \vec{k}_{r} \, , \vec{k})
\end{eqnarray*}$$\eqno{(3.24.2)}$$
with $ \vec{k}=\vec{k}_{1}+ \ldots + \vec{k}_{r} $  and 
$$
|\vec{k}_{1}+ \ldots + \vec{k}_{r}| \le d \quad \mbox{and} \quad \sum_{i=1}^{r} \epsilon^{1}(\vec{k}_{i} )\le c \, . 
\eqno{(3.24.3)}
$$
we have
$$
H_{\vec{v}\omega} \mbox{ when restricted to }{\cal H}_{\omega}^{c,d,'} \ge 0
\eqno{(3.24.4)}
$$

\textbf{Remark 2}
Due to the corrections $O(N^{-1})$ we do not know whether ${\cal H}_{\omega}^{c,d,'}$ in the above remark 1 comprises 
the whole of ${\cal H}_{\omega}^{c,d}$, although we conjecture that to be so. In particular, nothing can be said about 
the canonical free energy restricted to the subspace (3.24.1) in the thermodynamic limit. For 
other conjectures associated to the (supposedly subadditive) excitation spectrum of Bose gases, see (\cite{CDZ}, (1.10) 
et seq.), and 
(\cite{CDZ}, Appendix B). On ${\cal H}_{\omega}^{c,d,'}$ the general model of independent elementary excitations 
of \cite{SeW} is fully justified. We believe that a qualitatively similar assertion applies to a wide variety of models in 
condensed matter physics.

\textbf{Remark 3}
(3.12) corresponds to the original Landau assumption (1.1). The restriction on energy and momentum 
necessary in Proposition 1 seems to confirm Kadanoff's remarks cited in the introduction.

\subsection{The Huang-Yang-Luttinger (HYL) model}

Our second example is the Huang-Yang-Luttinger (HYL) model \cite{HYL}, whose pressuure was rigorously derived by van den 
Berg, Dorlas, Lewis and Pul\'{e} in \cite{BDLP}, where it is also proved that it exhibits BEC. Let
\begin{eqnarray*}
H_{\Lambda}^{0} \equiv \sum_{\vec{k} \in S_{\Lambda}^{d=3}} \frac{(\vec{k})^{2}n_{\vec{k}}}{2}
\end{eqnarray*}$$\eqno{(3.25)}$$
denote the free Hamiltonian, and
\begin{eqnarray*}
H_{\Lambda}^{HYL} \equiv H_{\Lambda}^{0} + \frac{\widetilde{a}(2(N^{op})^{2}-\sum_{\vec{k}}n_{\vec{k}}^{2})}{2V} \, , 
\end{eqnarray*}$$\eqno{(3.26)}$$
where $n_{\vec{k}}=a_{\vec{k}}^{\dag}a_{\vec{k}}$ denotes the number operator for the $\vec{k}$ -th mode, $\widetilde{a}$ is a positive 
number and $V=L^{3}$ denotes the volume of the cubic region and $N^{op}$ is the number operator (2.26). We have
\begin{eqnarray*}
a_{\vec{k}} = \frac{\int d\vec{x} \exp(-i\vec{k}\cdot\vec{x}) \Psi(\vec{x})}{\sqrt(V)} \, , 
\end{eqnarray*}$$\eqno{(3.27)}$$
while, in terms of the basic destruction operator $\Psi$ (see (2.27)) the group of Galilean transformations is given by (2.29). The free Hamiltonian (3.25) may be written \cite{AW}
$$
H_{\Lambda}^{0} = \int_{V} d\vec{x} \; \nabla \Psi^{\dag}(\vec{x}) \cdot \nabla \Psi(\vec{x}) \, , 
\eqno{(3.28)}
$$
which transforms under (2.29) to
$$
H_{\Lambda} \to H_{\Lambda}^{0} + \vec{v} \cdot \vec{P}_{\Lambda} +\frac{N(\vec{v})^{2}}{2} \, , 
\eqno{(3.29)}
$$
with
$$
\vec{P}_{\Lambda}= \frac{1}{2} \int {\rm d} \vec{x} \; [\Psi^{\dag}(\vec{x})(\nabla\Psi)(\vec{x})-(\nabla\Psi^{\dag}(\vec{x})\Psi(\vec{x})]
\eqno{(3.30)}
$$
the momentum operator; by (3.27), $a_{\vec{k}}^{\dag} \to a_{\vec{k}+\vec{v}}^{\dag}$, $a_{\vec{k}} \to a_{\vec{k}+\vec{v}}$ 
under (2.29.2) and thus $\sum_{\vec{k} \in S_{\Lambda}^{3}}n_{\vec{k}}^{2}$ remains invariant under (2.29.2) (recall our 
choice (2.13.3)). Therefore (3.29) becomes
\begin{eqnarray*}
H_{\Lambda} \to H_{\Lambda} + \vec{v} \cdot \vec{P}_{\Lambda} + \frac{N(\vec{v})^{2}}{2} \, , 
\end{eqnarray*}$$\eqno{(3.31)}$$
where we omit the suffix HYL from now on. By (\cite{Hu}, Section~10.5), the correspondence
$$
\widetilde{a} = 8 \pi a
\eqno{(3.32)}
$$
is valid, where $a > 0$ is the scattering length of the given repulsive potential, which is approximated by a poitwise (delta) 
repulsion in each order $j$ of perturbation theory (and only in a fixed order, since, globally, such a potential is not
stable in 
the sense of (2.12) --- see \cite{AGHH} --- it exhibits, however, nontrivial scattering \cite{AGHH}, in spite of contrary assertions 
in many books). In first order, we obtain (2.18.3),(2.18.4), with the correspondence (3.32), see (\cite{Hu}, (10.124) and (A36)).

Since $H_{\Lambda}$ depends only on the number operators and not on the creation and destruction operators individually, 
we may work on the $N$- particle Fock space ${\cal F}({\cal H}_{\Lambda})$, the symmetrized $N$- fold tensor product of 
${\cal H}_{\Lambda}=L^{2}(\Lambda)$ (with periodic b.c.). The ground state energy $E_{N,L}^{0}$ of $H_{\Lambda}=H_{N,L}$, 
given by (3.26), obtains upon setting $n_{\vec{0}}=N$, whereby
$$
E_{N,L}^{0} = \frac{\widetilde{a}(2N^{2}-N^{2})}{2V}= \frac{\widetilde{a}N^{2}}{2V}
$$
and the ground state energy per unit volume in the limit $N,L \to \infty$ is
$$
e_{0}(\rho) = E_{N,L}^{0}/V = \widetilde{a}\rho^{2}/2 = 4 \pi a \rho^{2}
\eqno{(3.33)}
$$
by (3.32). By (3.31), (3.33) and Definition 1 we arrive at the following result.

\textbf{Lemma 3} The HYL Hamiltonian (3.26) is an effective Hamiltonian to order $k=1$ for the dilute Boson system.

One reason that it is believed that $H_{\Lambda}$ is a good approximation (to order one) for the dilute system, besides being an 
effective Hamiltonian to order one, is that it may be regarded as a correction to the mean-field Hamiltonian
\[
H_{\Lambda}^{mf} \equiv H_{\Lambda}^{0} + \frac{\widetilde{a}N^{2}}{V}
\]
by incorporating a \emph{local} repulsion. Indeed, the mean-field Hamiltonian would also be an effective Hamiltonian to order one 
for the dilute Boson system (with a slightly different choice of parameters), but it does not exhibit superfluidity (in the Landau sense) 
for the simple reason that
\begin{eqnarray*}
H_{\Lambda}^{mf} - E_{\Lambda}^{mf,0} = H_{\Lambda}^{0} = H_{\Lambda}^{0} - E_{\Lambda}^{0,0} \, , 
\end{eqnarray*}$$\eqno{(3.34)}$$
where 
$$
E_{\Lambda}^{mf,0} = \frac{\widetilde{a}N^{2}}{V} = \widetilde{a} \rho^{2} V
\eqno{(3.35)}
$$
and $E_{\Lambda}^{0,0} = 0$ are the ground state energies of the mean field and the free Hamiltonian. Thus the eigenvalues and/or 
elementary excitations of $H_{\Lambda}^{mf}-E_{\Lambda}^{mf,0}$ are the same as those of the free Bose gas, which is not a 
Landau superfluid. The same remark in somewhat different language was made in (\cite{Ver}, pg.~85), where it is also remarked 
(pp.~82-84) that, for the mean-field model, however, the canonical and grand-canonical states coincide, in contrast to the free Bose gas. 
It may be added that, due to (3.35), $e_{0}(\rho) = \widetilde{a} \rho^{2}$ and, therefore, the compressibility is nonzero: $e_{0}^{''}(\rho) 
= 2 \widetilde{a} > 0$. The formula $c=(\frac{1}{\rho e_{0}^{''}(\rho)})^{1/2}$ for the sound velocity (see, e.g., \cite{WSi}) is not, however, true 
for the mean field model, because the latter is zero as in the case of the free Boson gas by (3.34).

Note that the Girardeau model exhibits superfluidity but no BEC by \cite{Len}. By the above, the mean field model exhibits BEC (see \cite{Ver}, pg. 85), but no superfluidity. Thus BEC and superfluidity are independent concepts, although, in a more realistic system with BEC, it may well be that the condensate may be identified with the superfluid part, as often intuitively assumed.

We now consider the Hamiltonian $H_{\vec{v},N,L}$ given by (2.19.1). Our main result for the HYL model is the following one.

\textbf{Proposition 2} If
$$
{\vec{v}_{lim} }^{2} < 2 \widetilde{a} \rho \, , 
\eqno{(3.36)}
$$
then the HYL model describes a NESS at $T=0$ with energy spectrum ,according to definition 3.2, unbounded from below. Moreover, defining in (2.30.2),
$$
\rho_{max} \equiv \rho - \frac{v_{lim}^{2}}{2\widetilde{a}} \, , 
\eqno{(3.37)}
$$
then it describes a $T=0$ superfluid, i.e., (2.33) of definition 4 holds.

\textbf{Proof} We have that
\[
H_{\vec{v},N,L} = \sum_{\vec{k} \in S_{\Lambda}^{3}} \frac{(\vec{k}-\vec{v})^{2}n_{\vec{k}}}{2}  
-\frac{Nv^{2}}{2} + \frac{\widetilde{a}(N^{2}-\sum_{\vec{k}}n_{\vec{k}}^{2})}{2V}
\]
(with $\vec{v} = \vec{v}_{\vec{n}_{L},L}$ defined by (2.13.2). The lowest eigenvalue $-\frac{Nv^{2}}{2}$ of
$H_{\vec{v},N,L}$ obtains for $n_{\vec{v}}=N$. In general, the energy is minimized by choosing, first, $n_{\vec{0}}=N-n$ together with 
$n_{\vec{v}}=n$, and then finding the minimum with respect to $n$. Indeed, a splitting of 
$N=n_{\vec{0}}+n_{\vec{k}_{1}}+ \ldots + n_{\vec{k}_{l}}$, with $n_{\vec{0}}=N-n$ and $\sum_{i=1}^{l}n_{\vec{k}_{i}}=n$ with
$l > 1$ entails a relative increase in energy of $\frac{2\widetilde{a}\sum_{\vec{k} \ne \vec{l}}n_{\vec{k}}n_{\vec{l}}}{V}$.

The corresponding eigenvalues $E_{\vec{v},N,L}$ of $H_{\vec{v},N,L}$ are
\begin{eqnarray*}
E_{\vec{v},N,L} = -\frac{nv^{2}}{2} + \frac{\widetilde{a}(Nn-n^{2})}{V}\\   
= -\frac{nv^{2}}{2} + (\widetilde{a} \rho n - \frac{\widetilde{a}n^{2}}{V}) \, . 
\end{eqnarray*}$$\eqno{(3.38)}$$
We now assume (3.36). The function
\[
f_{\vec{v}}(n) \equiv (\widetilde{a} \rho -v^{2}/2)n -\frac{\widetilde{a}n^{2}}{V}
\]
has a maximum at
\[
n_{max} \equiv (\frac{\widetilde{a}\rho-v^{2}/2}{2\widetilde{a}})V = N/2 - \frac{v^{2}V}{4\widetilde{a}} \, , 
\]
and minima at $n_{min}^{1}=0$, together with
\begin{eqnarray*}
n_{min}^{2} = \frac{V(\widetilde{a}\rho-v^{2}/2)}{\widetilde{a}} = N - \frac{v^{2}V}{2\widetilde{a}}
\end{eqnarray*}$$\eqno{(3.39)}$$
(we ignore the corrections due to the fact that these numbers are integers, which are easily shown to be negligible for large 
enough $V,N$). For
$$
n \le n_{min}^{2}
\eqno{(3.40)}
$$
the eigenvalues are positive: this assertion is, of course, always true for $\vec{v}=\vec{0}$ as it should be. By (3.39), (3.40), this yields the last assertion of the proposition.

Let, now, $k$ be an integer such that
$$
V(\rho - \frac{v^{2}}{2\widetilde{a}}) + k \le N
\eqno{(3.41.1)}
$$
or
$$
(\rho - \frac{v^{2}}{2\widetilde{a}}) + k/V \le \rho \, , 
\eqno{(3.41.2)}
$$
which always holds in the thermodynamic limit, for any fixed integer $k > 0$. Note that for $\vec{v} = \vec{0}$ the density 
interval (3.41) becomes empty, as it should; for $\vec{v} \ne \vec{0}$,
\[
f_{\vec{v}}(V(\rho - \frac{v^{2}}{2\widetilde{a}})+k) = -k(\widetilde{a}\rho -v^{2}/2) - \widetilde{a}/V
\]
yielding, by Corollary 1, the first assertion of the proposition; moreover, since the integer $k$ is arbitrary, the energy spectrum in the sense of definition 3.2 is unbounded from below. q.e.d.

\textbf{Remark 4}
 Proposition 2 may be regarded as a precise statement corresponding to Nozi\'{e}res' intuition, mentioned 
in the introduction \cite{Noz}: under (3.36), with $\rho_{max}$ given by (3.37), a fragmentation of the condensate costs a 
macroscopic amount of energy, rendering it stable. Further, (3.38) for small $n$ and $\vec{v}=\vec{0}$, exhibits his conjectured 
macroscopic stability, when $n=n_{\vec{k}}$ with~$\vec{k}$ arbitrarily close to the zero vector.

\section{Conclusion}

We have introduced a framework characterizing (Landau) superfluids as non-equilibrium stationary states (NESS) and shown its applicability to two models - the Girardeau model (and very probably the Lieb-Liniger model, although the computations have not been considered in detail) and the Huang-Yang-Luttinger (HYL) model. Stationarity comes about naturally from the physical input. 

Since local stability conditions are not expected to be valid for NESS (because, under certain conditions, they imply that the state is 
an equilibrium state (\cite{HKTP}\cite{BKR}) - this fact was actually proved by Ogata \cite{Ogata} for a special model -, it is natural to characterize a superfluid at $T=0$ by  certain metastability conditions (Definition 4), which physically express that there should be a sufficiently small 
energy-momentum transfer between 
the particles of the fluid and the surroundings (e.g., the pipe). They have been verified for the models of Section~3. In the case of the second 
model (HYL model) in Section~3, the metastability condition is directly related to Nozi\`{e}res' conjecture \cite{Noz} that it is the repulsive 
interaction which interdicts the fragmentation of the condensate, thereby assuring its quantum coherence, with the wave-function becoming 
a macroscopic observable by the condition of ODLRO.

The instability responsible for the state becoming a non-equilibrium state may be expected to be due to certain excitations inherent to the 
system, when their energy and/or momenta exceed certain values. Although excitations such as the umklapp excitations, imparting large 
momenta and low energy, may be special to singular repulsive one-dimensional interactions, the remarkable result of Seiringer and 
Yin \cite{SeYin}, showing that the Lieb-Liniger model is a suitable limit of dilute Bosons in three dimensions, indicates that a qualitatively 
similar picture may take place for more realistic systems. In general, for realistic systems, it is conjectured that that the instability in the case 
of a rotating bucket is caused by vortices (see, e.g.~\cite{MaRo}, pg.~235), which, for systems with nonzero density, have been seen to exist 
in the free Boson gas in \cite{LPu}. For translational superfluids, the instability is conjectured to be caused by rotons, which arise as local 
minima of the infimum of the excitation spectrum, see \cite{MaRo}, pg.~246. These should also take place only for sufficiently large momentum, 
as a consequence of the properties of the liquid structure factor, if one assumes the Feynman variational wave function, see \cite{MaRo}, pg.~256. 
For the Feynman variational wave function, see \cite{CDZ} and \cite{WSi}. In both models of section 3 the ultimate cause of instability is seen to be the \textbf{local} repulsion (or a caricature thereof).

Unfortunately, the HYL model of Section~3 displays a gap and is, therefore, an unrealistic model of a Boson superfluid. It has been conjectured, 
however, that that the gap disappears in second order (\cite{Hu}, pg.~330), and it would be very nice to prove this, together with the other 
properties characterizing a superfluid. The methods of \cite{Sei}, \cite{LSYng} and \cite{CDZ} should be relevant here. This would provide a theory of superfluidity at the level of effective Hamiltonians, which would be very interesting, with applications to real systems, see \cite{GK} and \cite{CHSR}, but still very far from the full theory of dilute Boson systems at order $j=2$. Even the latter would be, as Leggett remarked about Bogoliubov's theory \cite{Leg1}, ``extremely suggestive but rather far from real-life Helium II'', because the 
latter is a strongly correlated system, for which, in addition, the attractive part of the interaction is expected to play a significant role.

\textbf{Acknowledgements} I am grateful to Geoffrey Sewell for having stimulated my interest in this subject, to him and Christian J\"{a}kel for helpful remarks. I am very grateful to an anonimous referee for pointing out that the simple but important fact that the isotony property does not hold in general, in the case of periodic boundary conditions, as well as for further important constructive criticism.
\bibliography{minhabiblio-t}
\bibliographystyle{alpha}
\end{document}